\documentclass{article}
\usepackage{graphicx} 
\usepackage[margin=1.25in]{geometry}

\newcommand{\ph}[1]{\medbreak \noindent {\bf #1}}

\title{A Proposed End-To-End Principle \protect\\ for Data Commons}
\author{Robert L Grossman \protect\\
Center for Translational Data Science \protect\\
University of Chicago
} 
\date{June 15, 2018} 

\begin{document}

\maketitle

\begin{abstract}
A data commons brings together (or co-locates) data with cloud computing infrastructure and commonly used software services, tools and applications for managing, analyzing and sharing data to create an interoperable resource for a research community.  We introduce an architectural design principle for data commons called the narrow middle architecture that is broadly based upon the end-to-end argument in systems design.  We also discuss important core services for data commons and the role of standards.    
\end{abstract}

\ph{Introduction.} There are now several efforts that are building data commons for biomedical data, including the NCI Genomic Data Commons (GDC), the NIH Data Commons Pilot Consortium (DCPPC), and the NCI Cancer Research Commons (CRDC). I like to think of a data commons in the following way: a database organizes data for a project; a data warehouse organizes data for an organization; and a data commons organizes data for a field or discipline.

More formally, a data commons brings together (or co-locates) data with cloud computing infrastructure and commonly used software services, tools and applications for managing, analyzing and sharing data to create an interoperable resource for a research community \cite{grossman2016case}.

In a Medium post by Josh Denny, David Glazer, Benedict Paten, Anthony Philippakis and myself \cite{paten2017biosphere}, we proposed four governing principles for building data commons and related data platforms:

(1) modular, composed of functional components with well-specified interfaces; (2) community-driven, created by many groups to foster a diversity of ideas; (3) open, developed under open-source licenses that enable extensibility and reuse, with users able to add custom proprietary modules as needed; and (4) standards-based, consistent with standards developed by coalitions such as the Global Alliance for Genomics and Health (GA4GH). 


In this article, I would like to discuss principles (1) and (4) in more detail. In particular, I would like to discuss: how many modular components with well-specified interfaces are needed for a data commons and when should standards for these components be developed?

I want to emphasize that the opinions below are mine alone, and the community of data commons developers is still actively debating these and other principles. In particular, these opinions do not necessarily reflect the views of the projects that I am working on.

Before we begin, let us review two principles that are important when building distributed applications that support a community of users.

\ph{The end-to-end design principle.} The end-to-end principle \cite{saltzer1984end} is one of the design principles underlying the internet. It states that network features should be implemented as close to the end points of the network --- the applications --- as possible. This is commonly expressed by describing the system as a ``dumb'' network with ``smart'' applications at the endpoint. What this means in practice is that as new innovative applications are developed for creating new types of content (one ``end'') and for giving viewers access to the content (the other ``end''), the intermediate routers and switches do not need to be updated. A good example is provided by streaming media which is a new type of content that required new types of applications but did not require any changes to the core routers and switches that support the internet. Lawrence Lessig and Robert W. McChesney argue that the end-to-end principle is what has made the Internet such a success, since ``All of the intelligence is held by producers and users, not the networks that connect them \cite{lessig2006no-tolls}.''

\ph{The principle of avoiding premature standardization.} I find it helpful to think of standardization along a temporal spectrum. One end of the spectrum is when well accepted interfaces and specifications are standardized. The other end of the spectrum is when an organization or group develops a new standard to encourage the emergence of a new community. The difference is how many implementations are available, how long have they been used, and how widely adopted they are. The critical question is when in the lifecycle of a technology is the appropriate time to standardize? If too early, innovation is put at risk or the standard is likely to be ignored; if too late, the overall benefit is less.

We now discuss two possible architectures for a data commons: a standards-driven architecture and a narrow-middle architecture.

\ph{Standards driven architecture.}  With a standard driven architecture, a list of standards is developed at the beginning of the project, and the goal is to develop software consistent with the standards. This works well with fields that are relatively mature. The challenge with data commons is that the field is moving ahead quickly and the technologies for importing, curating, and integrating data are still maturing, as are the processes for exploring, analyzing, and collaborating with data.

\begin{figure}[ht]
\centering
\includegraphics[scale=0.32]{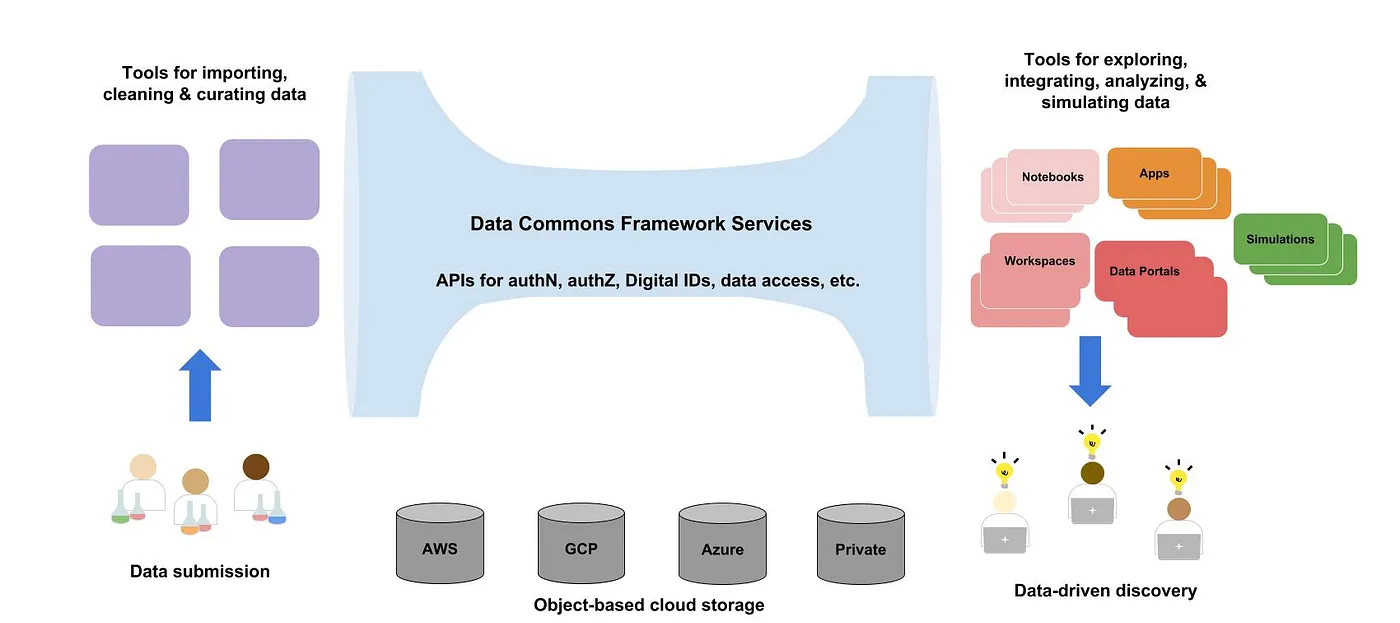}
\caption{The narrow middle architecture}
\label{fig:narrow-middle}
\end{figure}

\ph{Narrow middle architecture.} With what you might call a narrow middle architecture, the fewest possible services are used for the core of the system (``the narrow middle'') and only these services are standardized. See Figure~\ref{fig:narrow-middle}.  This is an example of an end-to-end design. With this approach, the importing, curation and integration services for getting data into the commons (one ``end'') and the data exploration, analysis and collaboration services for getting knowledge out of the commons (the other ``end'') are not standardized, but instead are the focus of different competing and innovative efforts until the community begins to recognize and adopt approaches that seem to be most effective.

\ph{Core services.} A good question to ask is: what are the minimum services that would enable the development and operation of a data commons? I argue that only four are needed:

\begin{enumerate}

\item  Permanent Digital IDs, which given a globally unique ID (GUID) return the one or more locations that contain the corresponding data object. We assume that the underlying data objects comprising datasets of interest are stored in public and private clouds that are accessible through S3-compatible APIs.

\item A metadata service, which given a GUID returns a core set of metadata.

\item Authentication and authorization services, which enable the authentication of users and also can determine whether a given authenticated user is authorized to see a particular dataset. In this category, I also include the associated security and compliance services.

\item  Data model services for defining a data model, and for validating, submitting and querying data with respect to this data model.

\end{enumerate}

Given these four services, a data commons can be developed. For example, the GDC developed a graphQL-based API over these services based upon an extensible data model. A variety of third party tools have also been built over this API, creating the beginning of an ecosystem. Several other data commons have also been built using this approach, including the BloodPAC Data Commons.

The GDC serves over 100,000 unique users a year, a data portal, a data exploration portal, a data submission portal, several million API calls a day and multiple petabytes of data over these core services. Note that the GDC does not currently use a separate metadata service, but returns metadata through its API, but could easily create a simple metadata service over its current API.

When the GDC was developed during 2014--2016, Jupyter notebooks were not very common and supporting them was not something we ever thought about, but given the GDC’s use of a narrow middle architecture and its open APIs, no changes were required to support Jupyter notebooks as they grew more popular, and now it is a very common way to analyze data from the GDC.

It is sometimes helpful to divide a project’s data into the project’s object data, which is managed with 1) -- 3), and the rest of the project's data, which includes its structured data, unstructured data, data schemas, ontologies, etc. There is no agreed upon name for the latter, so we will simply call it (in aggregate) the project's core data. The project's core data is managed with 4). Note that services 1) -- 3) are enough to make a project’s object data findable, accessible, interoperable, and reusable or FAIR \cite{wilkinson2016fair}, but it is currently challenging to do the same for the project's core data, given the lack of technology consensus in this area. Individual projects and individual commons will each make their own choices, and some of these will win out over time. In many cases, a project’s object data is 1000x or more larger than its core data. Because of this difference in size, it is quite practical to support and experiment with multiple copies of a project's core data, something that is not possible with a project’s object data.

\ph{Two recommendations.} I have two recommendations: The first recommendation is when a community stands up a data common to agree quickly on services 1) -- 3). The second recommendation is to hold off standardizing on the other services used for developing data commons, including data model services, but instead collaborate and compete on: a) applications for analyzing and exploring data in the commons (one end of the narrow middle architecture); and, b) applications for importing, cleaning and curating data in the commons (the other end). Over time, de facto standards and practices will emerge.

\end{document}